\documentclass[preprint,amsmath,amssymb,aps]{revtex4-1}
\usepackage{graphicx}
\usepackage[english]{babel}
\usepackage{amsmath}
\usepackage{booktabs}
\usepackage{subcaption}
\usepackage{multirow}
\usepackage{float}
\usepackage{lineno}\usepackage{dcolumn}
\usepackage{bm}

\begin{document}

\title{Fokker-Planck model for binary mixtures}

\author{Samarth Agrawal}
\affiliation{Engineering Mechanics Unit,\\
Jawaharlal Nehru Centre for Advanced Scientific Research, Jakkur, Bangalore  560064, India}
\author{S. K. Singh}
\affiliation{Engineering Mechanics Unit,\\
Jawaharlal Nehru Centre for Advanced Scientific Research, Jakkur, Bangalore  560064, India}
\author{Santosh Ansumali}
\thanks{email for correspondence: ansumali@jncasr.ac.in}
\affiliation{Engineering Mechanics Unit,\\
Jawaharlal Nehru Centre for Advanced Scientific Research, Jakkur, Bangalore  560064, India}
\date{\today}

\begin{abstract}
In dilute gas kinetic theory, model collision dynamics such as Bhatnagar-Gross-Krook (BGK) model \citep{bhatnagar1954model} is often used to get better insight and numerical modeling.
BGK model and its variants  assumes that highly nonlinear collision term can be replaced by a simple relaxation dynamics towards  Maxwell-Boltzmann distribution. \citet{lebowitz1960nonequilibrium}, 
proposed an alternative framework for collision  model, known as Fokker-Planck model, where the relaxation of an arbitrary distribution towards  Maxwell-Boltzmann distribution is
  modelled as a  drift-diffusion process in velocity space.  
  In the present manuscript, we extend the single component Fokker-Planck model to a  binary gas mixture  
  model over a large range of Schmidt numbers. We   prove that this  mixture models satisfy the necessary conservation laws and the $H-$theorem. 
\end{abstract}

\maketitle

\section{Introduction}

Fluid dynamics for gases at the continuum scale, i.e, when the system size is much larger than the mean free path, is well described by the Navier-Stokes-Fourier dynamics. In this description, one assumes that the system 
is locally close to thermodynamic equilibrium. A more general description of dilute gas dynamics is provided by the Boltzmann equation which describes the dynamics even far from away from the thermodynamic
equilibrium and at all Knudsen numbers, defined as $Kn = \lambda/L$, where $\lambda$ is the mean free path and $L$ the characteristic length of the system \citep{chapman1970mathematical, grad1949kinetic}. In 
the kinetic theory of gases, as developed by Boltzmann and Maxwell, one assumes gases to be composed of structureless point particle and provides a statistical description of the motion of particles in 
terms of the single particle distribution function. The Boltzmann equation is the time evolution equation for the distribution function where time evolution is represented as a sequence of 
free flight and binary collision described by an integro-differential term. Given the complex non-linear integro-differential form of the collision term, there is a long history  of representing the collision term
via model dynamics in the kinetic
theory. Perhaps the most famous and widely used model is the Bhatnagar-Gross-Krook (BGK) model \citep{bhatnagar1954model}, where one assumes
that complex collision term can be replaced by a relaxation dynamics towards equilibrium distribution of Maxwell-Boltzmann. BGK  model has correct hydrodynamic limit of Navier-Stokes-Fourier
dynamics and thermodynamic consistency of the Boltzmann equation as described by $H$-theorem. However, as the model describes the relaxation towards Maxwell-Boltzmann by a single relaxation time, all 
non-conserved moments such as  stress and heat flux, relax at the same rate. Thus, the BGK model is incapable of accurately model  all the transport coefficients and thus predicts 
Prandtl number (ratio of the thermal and momentum diffusion time) of dilute gases as unity in place of $2/3$. More sophisticated models such as the ellipsoidal statistical BGK (ES-BGK) model do not 
have such defects \citep{holway1966new}. Constructing model collision dynamics of relaxation type, which preserves both hydrodynamic and thermodynamic consistency is well understood for both single 
component gas and multi-component gas mixtures \citep{gorban1994general}.

\citet{lebowitz1960nonequilibrium}, proposed an alternative Fokker-Planck framework for collision  model where the relaxation of an arbitrary distribution towards  Maxwell-Boltzmann distribution is
modeled as a drift-diffusion process in the velocity space.  Even though this model has a correct hydrodynamic limit (with Prandtl number $3/2$) and  satisfies the $H$-theorem, it
was rarely used for rarefied gas dynamics applications until recently. This model has seen a revived interest in the last few years owing to the fact that Fokker-Planck model has an equivalent Langevin
dynamics which can be efficiently discretized \citep{GJ14}.  Recently it was shown  that the Prandtl number defect can be cured without compromising on its thermodynamic consistency by two 
independent methodologies. \citet{GTJ11} introduced a generalized nonlinear Fokker-Planck model to correct the Prandtl number while \citet{singh2015fokker} showed that the Prandtl number 
can be tuned by changing the drift term in the Fokker-Planck model. These recent advances have successfully enabled the kinetic modelling of gases for boundary value problems pertaining to 
engineering applications via the Fokker-Planck approximation \citep{GTJ11, singh2016gaseous}. However, this is limited to the single 
component case and techniques dealing with gas mixtures haven’t attained the same level of sophistication. Recently, \citet{GJ12M} introduced a
generalized nonlinear Fokker-Planck model for gas mixtures that correctly describes the conservation of mass, momentum and energy and the transfer between 
the components as well and also managed to recover the relevant transport coefficients. However, as of now, there is no proof available for the thermodynamic
consistency of this model. In the present manuscript, we present an alternative approach based on quasi-equilibrium models to introduce a Fokker-Planck model for
binary mixtures and verify the veracity of this model through some basic simulations.

The manuscript is organized as follows. First, the distribution function and macroscopic variables are introduced following which we explain the Boltzmann equation and its basic properties for rarefied gases. In the next section, various approximations to the Boltzmann collisional kernel including the BGK approximation, quasi-equilibrium models and the Fokker-Planck approximation for hydrodynamics are revisited. A brief description of the Boltzmann equation for binary mixtures and quasi-equilibrium distribution functions for the same are described. Based on these ideas, we introduce two Fokker-Planck models for different Schmidt numbers. In the following sections, the transport coefficients for these models are calculated and the numerical solution algorithm of the resulting Fokker-Planck equations is presented. Finally we discuss the various benchmark problems that were used to validate these models.

\section{The distribution function and macroscopic variables}

The kinetic theory of gases provides a statistical description of the motion of molecules in terms of the distribution function, $f(\textbf{x}, \textbf{c}, t)$ which is the probability density of finding a particle in the phase space in the neighbourhood of the point $(\textbf{x}, \textbf{c})$ where $\textbf{x}$ and $\textbf{c}$ are the position and the molecular velocity respectively \citep{chapman1970mathematical}. Similarly, for an $\mathcal{N}-$component mixture the description is provided in terms of $f_i(\textbf{x}, \textbf{c}_i, t)d\textbf{x}d\textbf{c}_i$ which is the probability of finding a particle of the $i$th type in $(\textbf{x}, \textbf{x} + d\textbf{x})$, possessing a velocity in the range $(\textbf{c}_i, \textbf{c}_i + d\textbf{c}_i)$. The relevant macroscopic quantities can then be found by taking the appropriate moment of the distribution function. The component number density $n_i$ and the mixture number density $n$ are

\begin{align}\label{numberDen}
\begin{split}
n_{i} = \langle 1, f_{i} \rangle, \quad n = \sum_{i} n_{i},
\end{split}
\end{align}
where the summation is over all components and the $\langle\phi_1, \phi_2\rangle$ operator is

\begin{equation}
 \langle \phi_1(\textbf{c}_{i}), \phi_2(\textbf{c}_{i}) \rangle = \int _{-\infty}^{\infty} \phi_1(\textbf{c}_{i}) \phi_2(\textbf{c}_{i})d\textbf{c}_{i}.
\end{equation}
  A convention of explicit summation  over all components  is used for mixture quantities. As an example the mixture density is defined as $\rho = \sum_{i} \rho_i$, where the component mass density is defined as $\rho_i = m_in_i$ with $m_i$ being the mass of each particle of $i$th component. Similarly, in $D$ dimensions, the momentum $\rho\textbf{u}$, the energy $E$ and the temperature $T$ of the mixture are defined as
\begin{align}
\begin{split}
\rho\textbf{u} = \sum_{i} \langle m_{i}\mathbf{c}_{i}, f_{i} \rangle,\quad E = \sum_{i} \left \langle \frac{m_{i}{c}_{i}^2}{2}, f_{i} \right \rangle, \quad \frac{D}{2}nk_BT = \sum_{i} \left \langle \frac{m_{i}(\mathbf{c}_{i} - \textbf{u})^2}{2}, f_{i} \right \rangle,
\end{split}
\end{align}
where $k_B$ is the Boltzmann constant. Similar to  the single component case, the component velocity $\textbf{u}_{i}$ and component temperature $T_{i}$ are
\begin{align}
\begin{split}
\rho_{i}\textbf{u}_{i} = \langle m_{i}\mathbf{c}_{i}, f_{i} \rangle,\quad \frac{D}{2}n_{i}k_BT_{i} = \left \langle \frac{m_{i}(\mathbf{c}_{i} - \textbf{u}_{i})^2}{2}, f_{i} \right \rangle.
\end{split}
\end{align}
Similarly the pressure $p$, the stress $\sigma_{\alpha\beta}$ and the heat flux $q_{\alpha}$ are

\begin{equation}
p = \sum_{i} n_ik_BT, \qquad \sigma_{\alpha\beta} = \sum_{i} \left\langle m_i\overline{\xi_{i\alpha} \xi_{i\beta}}, f_i \right\rangle, \qquad q_{\alpha} = \sum_{i} \left\langle m_i\frac{\xi_{i}^2}{2}\xi_{i\alpha}, f_i \right\rangle,
\end{equation}
where $\xi_{i\alpha} = c_{i \alpha} - u_{\alpha}$  and with $\overline{A_{\alpha\beta}}$ indicating the traceless part of the tensor. At equilibrium, the distribution function attains the Maxwell-Boltzmann distribution  form

\begin{equation}
   f^{{\rm MB}}_{i}\left(n_i, {\bf u}, T \right)  = n_{i}\left(\frac{m_{i}}{2 \pi k_B \,T } \right)^{D/2}
\exp{\left(-\frac{m_{i}}{2k_B T} (\mathbf{c_{i}} - \mathbf{u})^2 \right)}.
 \end{equation}
The component velocities and temperatures assume the value of their mixture counterparts while the stress and heat flux become zero at equilibrium, that is

\begin{equation}
  \textbf{u}_{i}[f^{\rm MB}]  =  \mathbf{u}, \quad T_{i}[f^{\rm MB}]  = T, \quad \sigma_{\alpha\beta}[f^{\rm MB}] = 0,     \quad q_{\alpha}[f^{\rm MB}] = 0.
\end{equation}
The kinetic theory of gases also extends the idea of entropy present in statistical mechanics to non-equilibrium situations. This is achieved via the $H-$function defined as

\begin{equation}
\label{hdefinition}
H = \sum_{i} \int d\textbf{c}_{i}(f_{i}\ln f_{i} - f_{i}).
\end{equation}
It can be shown that the $H-$function steadily decreases as the system progresses in time and at equilibrium attains a form similar to the Sackur-Tetrode expression of
entropy per unit mass in the thermodynamics \citep{chapman1970mathematical}, as $S_B = -k_B H[f^{\rm MB}]$, wherein
\begin{equation}
 H[f^{\rm MB}] = \sum_{i} n_i \left[\frac{D}{2}\log\frac{2\pi k_BT}{m_i} - \log n_i + \frac{D}{2} \right],
\end{equation}
 which shows that kinetic theory is consistent with features of statistical mechanics. This completes the description of the various relevant macroscopic quantities that are calculated 
 from the distribution function and their behaviour at equilibrium.

\section{The Boltzmann equation}
For the case of dilute gases,
the time evolution of the distribution function is described by the Boltzmann equation \citep{chapman1970mathematical}. For the single component dilute gas, the Boltzmann equation has the form

\begin{equation}
\partial _t f \left({\bf x}, {\bf c}, t\right)+ \partial _{c_{\alpha}} f\left({\bf x}, {\bf c}, t\right) = \Omega,
\end{equation}
where $\Omega$ is the term which accounts for the change due to collisions between particles. The Boltzmann collisional operator, $\Omega^{\rm B}$, quantifies the change in the distribution function from all possible binary collisions, and is expressed as

\begin{equation}
 \Omega^{\rm B} = \int \int \int \left( w^{\prime}f\left({\bf x}, {\bf c}^{\prime}, t\right) f\left({\bf x}, {\bf c}^{\prime}_1, t\right) -
 w\, f\left({\bf x}, {\bf c}, t\right) f\left({\bf x}, {\bf c}_1, t\right) \right){\rm d}\mathbf{c}_{1}{\rm d}\mathbf{c}^{\prime}{\rm d}\mathbf{c}^{\prime}_{1},
\end{equation}
where $w \equiv w(\mathbf{c}^{\prime}, \mathbf{c}^{\prime}_{1}; \mathbf{c}, \mathbf{c}_{1})$ is the probability of the colliding pair to transition from the velocities $(\mathbf{c}, \mathbf{c}_{1})$ to $(\mathbf{c}^{\prime}, \mathbf{c}^{\prime}_{1})$ and vice-versa for $w^{\prime}$, and it can be shown that $w = w^{\prime}$ \citep{lifschitz1983physical}. The first term of the integrand represents the increase (gain) in the value of distribution function $f\left({\bf x}, {\bf c}, t\right)$ and similarly the second term represents decrease (loss). In order to satisfy the conservation of momentum and energy, the velocity pairs must satisfy

\begin{equation}
 \mathbf{c} + \mathbf{c}_{1} = \mathbf{c}^{\prime} + \mathbf{c}^{\prime}_{1}, \quad c^2 + c^{2}_1 = c^{\prime 2} + c^{\prime 2}_{1}.
\end{equation}
Therefore, by integrating over all possible $(\mathbf{c}_{1}, \mathbf{c}^{\prime}, \mathbf{c}^{\prime}_{1})$, the total change in $f(\mathbf{x}, \mathbf{c}, t)$ from
 collisions can be calculated. Further by considering appropriate moments and integrating over the velocity space, the dynamics of various macroscopic quantities can be derived \citep{grad1949kinetic,chapman1970mathematical}.

The Boltzmann equation is a highly complex integro-differential equation and hence does not lend itself to analysis even for simple boundary value problems. Therefore, approximations 
are made to the collisional term  to obtain a simplified description. A highly idealized, yet  quite effective model is  BGK-approximation \citep{bhatnagar1954model}, where the Boltzmann collisional operator is modelled as approach to the equilibrium distribution function. The BGK-collisional operator is

\begin{equation}
\Omega ^{\rm BGK} = \frac{1}{\tau_{\rm BGK}}\left(f^{\rm MB} - f \right),
\end{equation}
where $\tau_{\rm BGK}$ is the relaxation time. A variant of this approach is the ellipsoidal statistical BGK (ES-BGK) model \citep{holway1966new} which has the form

\begin{equation}
\Omega ^{\rm ES} = \frac{1}{\tau_{\rm ES}}\left(f^{\rm ES} - f \right),
\end{equation}
where $\tau_{\rm ES}$ is the relaxation time associated with this model and $f^{\rm ES}$ is the anisotropic Gaussian which has the form

\begin{equation}
  f^{\rm ES} = \frac{n}{\sqrt{{\rm det}[2\pi\lambda_{\alpha\beta}]}}\exp\left(-\frac{1}{2}\xi_{\alpha}\lambda_{\alpha\beta}^{-1}\xi_{\beta}\right),
\end{equation}
where ${\rm det}[]$ is the determinant and $\lambda_{\alpha\beta}$ is

\begin{equation}
 \lambda_{\alpha\beta} = \frac{k_B T}{m}\delta_{\alpha\beta} + b\frac{\sigma_{\alpha\beta}}{\rho},
\end{equation}
where the $b$ parameter is used to tune the Prandtl number as opposed to the BGK model where the Prandtl number is set to $1$. Another method of approximation is the Fokker-Planck operator \citep{lebowitz1960nonequilibrium}, wherein the approach to equilibrium is modelled as drift and diffusion dynamics

\begin{equation} \label{FPModel}
\Omega ^{\rm FP} =  \frac{1}{\tau_{{\rm FP}}} \partial _{c_{\alpha}} \left(\xi _{\alpha}f + \frac{k_BT}{m}\frac{\partial f}{\partial c_{\alpha}} \right) =  \frac{1}{\tau_{\rm FP}}\frac{k_BT}{m}\partial_{c_{\alpha}}\left(f \partial_{c_{\alpha}} \left(\ln f - \ln f^{\rm MB} \right) \right),
\end{equation}
which is essentially the diffusion dynamics in velocity space, with $\xi_{\alpha}$ acting as the drift coefficient, $k_BT/m$ assumes the role of diffusion coefficient and $\tau ^{-1}_{{\rm FP}}$ is the friction constant. It has been recently shown that this approximation is a useful alternative to methods such as DSMC for simulating moderately high Kn flows.

The Boltzmann equation and its aforementioned approximations are shown to have the following properties
\begin{enumerate}

\item {\bf Conservation Laws}: As binary collisions do not change mass, momentum or energy of the system, we have

\begin{equation}
\langle \Omega, \{m, m\textbf{c}, mc^2/2\} \rangle = \{0, \mathbf{0}, 0\},
\end{equation}
using this result and calculating appropriate moments of the Boltzmann equation, the conservation laws are

\begin{align}
\label{eq:conLaw}
\begin{split}
\partial_{t} \rho + \partial_{\alpha}\rho u_{\alpha} &= 0,\\
\partial_{t} \rho u_{\alpha} + \partial_{\beta}(\rho u_{\alpha}u_{\beta}+p \delta_{\alpha \beta})+\partial_{\beta} \sigma_{\alpha \beta}
&= 0, \\
\partial_{t}E+\partial_{\alpha}\left((E+p) u_{\alpha} +\sigma_{\alpha \gamma} u_{\gamma}\right)+\partial_{\alpha} q_{\alpha} &= 0,
\end{split}
\end{align}
which are in accordance the macroscopic laws of conservation.
\item {\bf Zero of the collision}:  When the collisions between particles do not affect the state of the system, it reaches a state of equilibrium 

\begin{equation}
\Omega = 0 \implies f = f^{\rm MB}.
\end{equation}
The converse is also true, i.e, when $f = f^{\rm MB}$ then $\Omega = 0$.

\item {\bf $H-$theorem}: The Boltzmann equation extends the idea of entropy to non-equilibrium situations. This is highlighted from the evolution of the $H-$function

\begin{equation}
\partial _t H + \partial _{\alpha} J_{\alpha}   = -\sigma^{S},
\end{equation}
where the $H-$function is $\left<(\text{ln}f - 1), f \right>$, $J_{\alpha}$ the entropy flux term and $\sigma^{S} = \left<\Omega, \text{ln}f \right>$ is the entropy generation term. The Boltzmann collisional operator, the BGK and Fokker-Planck approximations ensures that

\begin{equation}
\sigma^{S} \geq 0,
\end{equation}
entropy production is greater or equal to zero and hence the Boltzmann equation for rarefied gases is in accordance with the laws of thermodynamics. It is also noted that entropy production is zero at equilibrium, that is when $f = f^{\rm MB}$.

\end{enumerate}

This concisely summarizes the features of Boltzmann equation and its approximations for the single component case. In the following section we briefly explain the Boltzmann equation for binary mixtures, some well-known approximations and outline their important features.

\section{Kinetic modelling of binary mixtures}

\begin{figure}
\centering
\includegraphics[scale=0.4]{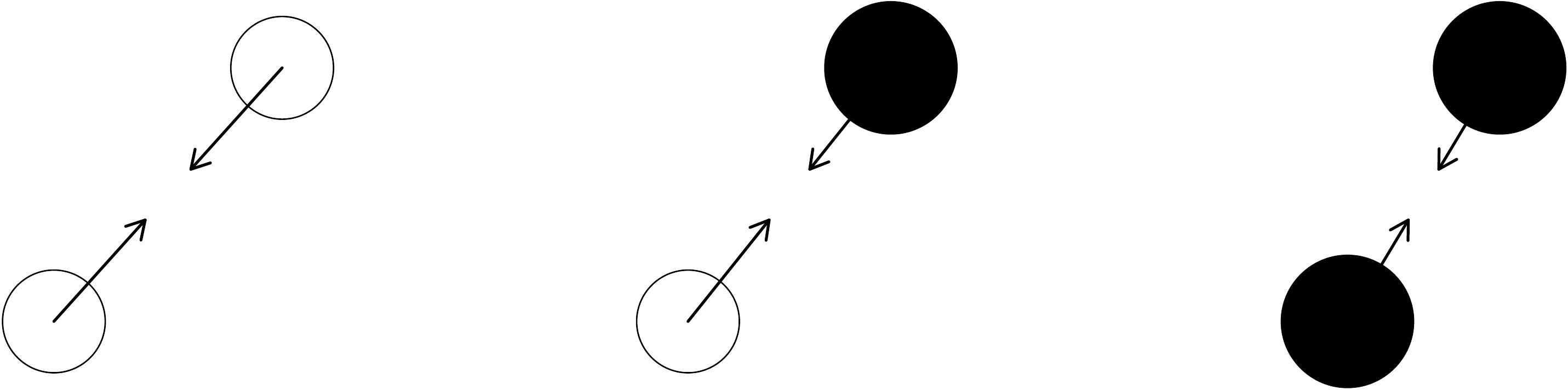}
\caption{The three types of collisional possibilities -- A-A, A-B and B-B}
\label{collisions}
\end{figure}

The dynamics of binary mixtures is fundamentally different from the single component case as the two components exchange momentum and energy through collisions, hence the Boltzmann equation for binary mixtures considers the different collisional possibilities, as schematically shown in Fig. \ref{collisions}. The changes in distribution of component A arises from A-A and A-B type collisions, and vice-versa for component B. As a result the mixture momentum and energy are conserved as opposed to their component wise counterparts \emph{i.e.} 
\begin{equation}
 m_{A}\mathbf{c}_{A} + m_{B}\mathbf{c}_{B} = m_{A}\mathbf{c}^{\prime}_{A} + m_{B}\mathbf{c}^{\prime}_{B}, \quad m_{A}c_{A}^2 + m_{B}c^{2}_{B} =  m_{A} c^{\prime 2}_{A} +  m_{B} c^{\prime 2}_{B},
\end{equation}
where $m_{A}$ and $m_{B}$ are the mass of $A$ type and $B$ type particles, respectively. Here $\mathbf{c}$  and $\mathbf{c^{\prime}}$ with subscript denote pre collision and post collision velocities of respective particles. The Boltzmann equation for binary mixtures is \citep{chapman1970mathematical}

\begin{align}
\begin{split}
\frac{\partial f_{A}}{\partial t} + c_{A\alpha}\frac{\partial f_{A}}{\partial x_{\alpha}} &= \Omega^{\textrm{B}}_{A} = \underbrace{\Omega^{\textrm{B}}(f_{A}, f_{A})}_{\textrm{Self-collision}} + \underbrace{\Omega^{\textrm{B}}(f_{A}, f_{B})}_{\textrm{Cross-collision}}, \\
\frac{\partial f_{B}}{\partial t} + c_{B\alpha}\frac{\partial f_{B}}{\partial x_{\alpha}} &= \Omega^{\textrm{B}}_{B} = \underbrace{\Omega^{\textrm{B}}(f_{B}, f_{B})}_{\textrm{Self-collision}} + \underbrace{\Omega^{\textrm{B}}(f_{B}, f_{A})}_{\textrm{Cross-collision}},
\end{split}
\end{align}
where the right-hand side of the equation is the change in distribution of the respective components arising from self collisions which is represented by $\Omega^{\textrm{B}}(f_{i}, f_{i})$ and $\Omega^{\textrm{B}}(f_{i}, f_{j})$ for cross collisions. Similar to the single component case, the evolution equation of various macroscopic variables can be derived using this equation. The collisional operator holds the following properties, which should ideally be satisfied by its approximations.

\begin{enumerate}
 \item {\bf Conservation Laws}: The mass of individual species as well as the total momentum and energy of the mixture are conserved as binary collisions do not contribute any change to these quantities, this is represented as

\begin{align} \label{binaryConstraints}
\langle \Omega^{\textrm{B}}_{i}, \{ m_{i}, \sum_{i=A,B} m_{i}\textbf{c}_{i}, \sum_{i=A,B} m_{i}\frac{c^{2}_{i}}{2} \} \rangle = \{ 0, \mathbf{0}, 0 \}
\end{align}
using which the conservation laws can be calculated similar to the single component case. However, the componentwise momentum and energy are not conserved as the two components exchange momentum and energy between themselves through cross-collisions (A-B type collisions). It is in fact these collisions that facilitate the relaxation of the component momentum and energy to the mixture momentum and energy \citep{H65}. The mixture variables adhere to the conservation laws as mentioned in Eq.(\ref{eq:conLaw}).

\item {\bf Equilibrium}: Similar to the single component case, the system reaches a state of statistical equilibrium. The distribution of any component $i$ at equilibrium is

\begin{equation}
   f^{\rm MB}_{i}  = n_{i}\,\left(\frac{m_{i}}{2 \pi k_B \,T } \right)^{3/2}
\exp{\left(-\frac{m_{i}}{2\, k_B T} (\textbf{c}_{i}-\textbf{u})^2 \right)}.
 \end{equation}
 The converse is also true, i.e, when $\Omega^{\rm B}_{i} = 0$ then the distribution function attains the form $f = f^{\rm MB}_{i}$.

\item {\bf $H-$theorem:} The Boltzmann collision kernel for binary mixtures, satisfies the $H-$theorem, that is $\sigma^{(s)} \geq 0$ similar to the single component case.

\item {\bf Indifferentiability principle:} The equations should adhere to the indifferentiability prinicple, i.e, the equation should converge to the single component case for $m_A = m_B$.

\end{enumerate}

\begin{figure}
\centering
\includegraphics[scale=1.25]{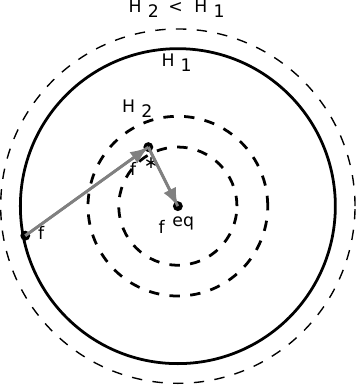}
\caption{Trajectory of the distribution function in a quasi-equilibrium model. The various concentric circles depict different levels of H}
\label{QE}
\end{figure}

Similar to the single component case, the corresponding BGK collision kernel for a binary mixture is

\begin{equation}
\Omega^{\rm BGK} = \frac{1}{\tau}(f^{\rm MB}_{i}(\rho_i, \textbf{u}, T) - f_{i}).
\end{equation}
The fundamental drawback with such a model is that there is only a single relaxation rate for all quantities whereas for the case of a binary mixture, there are two important time scales present in the system -- the rate of mass diffusion and the rate of momentum diffusion. The dimensionless parameter that is used to characterize these time scales is known as the Schmidt number and is defined as \citep{bergman2011fundamentals}

\begin{equation}
 {\rm Sc} = \frac{\text{viscous diffusion rate}}{\text{mass diffusion rate}} = \frac{\mu}{\rho D_{AB}},
\end{equation}
where $\mu$ is the viscosity, $\rho$ the density and $D_{AB}$ is the mass diffusion coefficient. For such an approximation, ${\rm Sc} = 1$ for all cases, and hence does not manage to accurately describe the system. Therefore, the collision kernel should be approximated in a manner capable of preserving these different time scales. Thus various approaches to correct this defect exist. In order to deal with multiple time scales, the basic idea of the fast - slow decomposition of motions near the quasi- equilibrium was introduced~\citep{gorban1994general,levermore1996moment}. In accordance with this idea, the relaxation to the equilibrium is modelled as a two-step process where `fast' relaxation happens from initial to quasi-equilibrium state and `slow'  relaxation happens from quasi-equilibrium state to  final equilibrium state. In the context of multiple time scales, the quasi-equilibrium models are a simple alternative \citep{levermore1996moment} to the BGK-approximation which can effectively incorporate multiple time scales of the system. The collision kernel for the quasi-equilibrium model is

\begin{equation}
 \Omega ^{\textrm{QE}}_{i} = \frac{1}{\tau _1}(f^{*}_{i}(M^{\textrm{quasi-slow}}, M^{\textrm{slow}}) - f_{i}) + \frac{1}{\tau _2}(f^{\textrm{MB}}_{i}(M^{\textrm{slow}}) - f^{*}_{i}(M^{\textrm{quasi-slow}}, M^{\textrm{slow}})),
\end{equation}
where $f^{*}_{i}(M^{\textrm{quasi-slow}}, M^{\textrm{slow}})$ is the quasi-equilibrium distribution function and is a function of the quasi-slow moments, $M^{\textrm{quasi-slow}}$ and the slow moments $M^{\textrm{slow}}$ \citep{levermore1996moment}. The idea is that the system moves towards a state of quasi-equilibrium where the quasi-slow moments relax first and then proceed towards equilibrium where the slow moments react, a visual description of the idea is presented in Fig. \ref{QE}. In accordance with the slow-fast dynamics that emerges from quasi-equilibrium models, two possible forms for the quasi-equilibrium distribution can be chosen -- for low ${\rm Sc}$ where mass diffusion occurs at higher rate as compared to momentum diffusion and vice versa for the high ${\rm Sc}$ case. For the first case, the physically relevant quasi-slow variables are

\begin{equation}
 M^{\textrm{quasi-slow}} = \{\rho_{i}, \rho_{i}\textbf{u}_{i}, n_{i}k_B T_{i}\},
\end{equation}
which imposes the following conditions on quasi-equilibrium distribution function $f^{*}_{i}$

\begin{align}
 \begin{split}
  \langle \{m_{i}, m_{i}\textbf{c}_{i}, m_{i}\frac{(\textbf{c}_{i} - \textbf{u}_{i})^2}{2} \}, f^{*}_{i} \rangle &= \{\rho _{i}, \rho _{i} \textbf{u}_{i}, \frac{3}{2}n_{i}k_B T_{i} \}.
  \end{split}
 \end{align}
By minimizing the $H-$ function as defined in Eq.(\ref{hdefinition}) under these constraints, the form of quasi-equilibrium for Low Schmidt limit, $f^{*(L)}_{i}$ is \citep{amak06}

\begin{equation} \label{lowQE}
 f^{*(L)}_{i} = n_{i}\left(\frac{m_{i}}{2\pi k_B T_{i}}\right)^{3/2}\exp{\left(-\frac{m_{i}(c_{i} - u_{i})^2}{2k_BT_{i}}\right)}.
\end{equation}
Similarly, for the second case where the momentum diffuses faster, the set of constraints under which the $H-$function is to be minimized are
\begin{align}
 \begin{split}
  \langle \{m_{i}, m_{i}\textbf{c}_{i}, \sum_{i=A,B} m_{i}\xi_{i\alpha}\xi_{i\beta}\}, f^{*}_{i} \rangle &= \{\rho _{i}, \rho _{i} \textbf{u}, n\theta_{\alpha\beta} \},
  \end{split}
 \end{align}
where
\begin{equation}
\theta _{\alpha\beta} = \frac{1}{n}\sum_{i=A,B}\left\langle m_{i}\xi_{i\alpha}\xi_{i\beta}, f_{i} \right\rangle.
\end{equation}
The quasi-equilibrium distribution function for high Schmidt limit $f^{*(H)}_{i}$ is\citep{amak06}

\begin{equation} \label{highQE}
 f^{*(H)}_{i} = n_{i}\left(\frac{m_{i}}{2\pi |\theta_{\alpha\beta}|} \right)^{\frac{3}{2}}\exp\left(\frac{-m_{i}\xi_{i\alpha}\theta^{-1}_{\alpha\beta}\xi_{i\beta}}{2}\right),
\end{equation}
where $|\theta_{\alpha\beta}|$ is the determinant.

These two distinct forms of quasi-equilibrium can be used to build two different collision kernels based on the Fokker-Planck approximation, which can solve for binary mixtures.

\section{Quasi-equilibrium models for Fokker-Planck formulation}

The Fokker-Planck approximation to the Boltzmann equation first introduced in Eq.(\ref{FPModel}) involves only a single time scale and therefore not well suited for modelling systems with multiple time scales. Hence, in order to extend the Fokker-Planck approximation for binary mixtures the concept of quasi-equilibrium models must be incorporated in a manner that correctly represents the multiple time scales present in the system and its approach to the equilibrium.

From Eq.(\ref{FPModel}), the Fokker-Planck approximation is
\begin{equation}
 \Omega^{\rm FP} = \frac{1}{\tau_{\rm FP}}\frac{k_BT}{m}\partial_{c_{\alpha}}\left(f \partial_{c_{\alpha}} \left(\ln f - \ln f^{\rm MB} \right) \right).
\end{equation}
This form of the Fokker-Planck model better illustrates the approach of the Maxwell-Boltzmann distribution, similar to the BGK model. In order to build a quasi-equilibrium like model with multiple time scales, we extend the FP model for it to have a similar form. Here, the approach to equilibria is defined as a two-step process wherein the first term represents a logarithmic approach to the quasi-equilibrium and the second to the equilibrium:
\begin{equation} \label{logFP}
 \Omega^{\rm FP}_{i} = \left(\frac{1}{\tau_{1}} - \frac{1}{\tau_{2}} \right)\mathbf{A}_{\alpha\beta}\partial_{c_{i\beta}}\left(f_i \partial_{c_{i\alpha}} \left(\ln f_i - \ln f_{i}^{*} \right) \right) + \frac{1}{\tau_{2}}\frac{k_BT}{m_{i}}\partial_{c_{i\alpha}}\left(f_i \partial_{c_{i\alpha}} \left(\ln f_i - \ln f_{i}^{\rm MB} \right) \right),
\end{equation}
where $\tau_1$ and $\tau_2$ are the characteristic time scales associated with the approach to quasi-equilibrium and equilibrium, and $\mathbf{A}_{\alpha \beta}$ is the diffusion coefficient that relaxes the system to the quasi-equilibrium state, for the low Schmidt dynamics, $\mathbf{A}_{\alpha \beta}$ can be chosen as  $\mathbf{A}_{\alpha \beta} = (k_{B} T_{i}/ m_{i}) \delta_{\alpha\beta} $. Using the form of $f^{*}_{i}$ presented in Eq.(\ref{lowQE}), the collision kernel for the low Schmidt limit is

\begin{equation}
\label{fp1}
\Omega_{i}^{\rm FP(L)} = \frac{1}{\tau_{1}}\partial_ {c_{i\alpha}} \left( (c_{i\alpha} - u_{i\alpha}) f_{i} +
\frac{k_{B} T_{i}}{m_{i}} \frac{\partial f_{i}}{\partial c_{i\alpha}}\right) +
\frac{1}{\tau_{2}}\partial_ {c_{i\alpha}} \left( (u_{i\alpha} - u_{\alpha}) f_{i} +
\frac{k_{B} \Delta T}{m_{i}} \frac{\partial f_{i}}{\partial c_{i\alpha}}\right),
\end{equation}
where $\Delta T = T - T_{i}$ is the difference in component and mixture temperatures. Similarly, for the high Schmidt dynamics, $\mathbf{A}_{\alpha \beta}$ is taken as  $\mathbf{A}_{\alpha \beta} = \theta_{\alpha \beta}$ and the collision kernel for the high Schmidt limit is

\begin{equation}
\label{fp2}
\Omega^{\rm FP(H)}_{i} = \frac{1}{\tau_{1}}\partial_ {c_{i\alpha}} \left( (c_{i\alpha} - u_{\alpha}) f_{i} +
\frac{ \theta_{\alpha \beta}}{m_{i}} \frac{\partial f_{i}}{\partial c_{i\beta}}\right) + \frac{1}{\tau_{2}}
\partial_ {c_{i\alpha}} \left(  \left( \frac{k_{B} T \delta_{\alpha \beta}}{m_{i}} - \frac{ \theta_{\alpha \beta}}{m_{i}} \right)\frac{\partial f_{i}}{\partial c_{i\beta}}\right),
\end{equation}
For this model to be considered canonical, it must satisfy the properties of collision as mentioned in section 4. By integrating over the velocity space $\textbf{c}_{i}$, it can be verified that the quasi-equilibrium FP model satisfies the constraints of Eq.(\ref{binaryConstraints}). The evolution equations for component mass, mixture momentum and energy are the same as the conservation laws mentioned in Eq.(\ref{eq:conLaw}). Furthermore, the component momentum and energy equations in relaxation form are

\begin{align}
\begin{split}
\label{componentrelax1}
 \left\langle \Omega^{\textrm{FP}(L)}_{i}, m_{i}c_{i\alpha} \right\rangle &= \frac{1}{\tau_{2}} \left( \rho_{i} u_{\alpha} - \rho_{i} u_{i\alpha}\right), \\
 \left\langle \Omega^{\textrm{FP}(L)}_{i}, \frac{m_{i}c_{i}^{2}}{2} \right\rangle &= \frac{1}{\tau_{2}} \left( \rho_{i} u_{i\alpha} (u_{\alpha} - u_{i\alpha}) + D k_{B} n_{i} (T - T_{i}) \right),
\end{split}
\end{align}
for the low Schmidt case. Similarly, for the high Schmidt case the relaxation equations for component momentum and energy are

\begin{align}
\begin{split}
\label{componentrelax2}
 \left\langle \Omega^{\textrm{FP}(H)}_{i}, m_{i}c_{i\alpha} \right\rangle &= \frac{1}{\tau_{1}} \left( \rho_{i} u_{\alpha} - \rho_{i} u_{i\alpha}\right), \\
 \left\langle \Omega^{\textrm{FP}(H)}_{i}, \frac{m_{i}c_{i}^{2}}{2} \right\rangle &= \frac{1}{\tau_{1}} \left( \rho_{i} u_{i\alpha} (u_{\alpha} - u_{i\alpha}) + D k_{B} n_{i} (T - T_{i}) \right).
\end{split}
\end{align}
 If $\tau_1 \leq \tau_2$,  the component velocities equilibrate faster in the second case than the first, which is as expected since the second model is applicable for high Schmidt regime wherein the viscous diffusion rate dominates the mass diffusion rate.

For the proposed model, the expression for entropy generation ($\sigma ^S$), is

\begin{align}
\begin{split}
\label{eq:H1}
 \sigma^{S}
= \frac{1}{\tau_{\textrm{eff}}} \sum_{i} - D n_{i}+ \int \frac{1}{f_{i}} \frac{\partial f_{i}}{\partial c_{i\alpha} }\mathbf{A}_{\alpha\beta}\frac{\partial f_{i}}{\partial c_{i\beta}} + \frac{1}{\tau_{2}} \sum_{i} - D n_{i} + \int \frac{k_{B} T}{m_{i}}\frac{1}{f_{i}} \frac{\partial f_{i}}{\partial c_{i\alpha} } \frac{\partial f_{i}}{\partial c_{i\alpha}}
\end{split}
\end{align}

where $\tau_{\textrm{eff}} = \tau_{2}\tau_{1}/(\tau_{2} - \tau_{1})$ and $\mathbf{A}_{\alpha\beta} = (k_BT/m_i) \delta_{\alpha\beta}$ for the low Schmidt case and $\mathbf{A}_{\alpha\beta} = \theta_{\alpha\beta}/m_i$ for the high Schmidt case. Following \citet{singh2015fokker}, Eq.(\ref{eq:H1}) can be rewritten as

\begin{align}
\sigma^{S} = \sum_{i} \underbrace{\int f_{i} \frac{\partial \ln \left ( f_{i}/ f^{*}_{i}\right)} {\partial c_{i\alpha}} \mathbf{A}_{\alpha\beta} \frac{\partial \ln \left( f_{i}/ f^{*}_{i}\right)}{\partial c_{i\beta}}
d{\bf c}_{i}}_{\rm positive} + \sum_{i} \underbrace{\int \frac{k_{B}T}{m_{i}}f_{i}\left(\frac{\partial \ln \left( f_{i}/ f^{*}_{i}\right)}{\partial c_{i\alpha}}\right)^{2} d{\bf c}_{i}}_{\rm positive},
\end{align}
which suggests that

\begin{equation}
\sigma^S  \geq 0, ~~~~~ \forall ~ \tau_{1} \leq \tau_{2}.
\end{equation}
Therefore, proposed model satisfies the $H-$theorem for $\tau_1 \leq \tau_2$.

An important condition for $\Omega^{\rm FP}$ to be considered valid is that the zero of collision must imply that the distribution function has attained a Maxwell-Boltzmann form. For the present model the zero of collision, i.e, $\Omega_{i}^{\rm FP} = 0$ implies

\begin{align}
\begin{split}
 \left \langle  \Omega^{\textrm{FP}(1)}_{i}, m_{i}c_{i\alpha} \right \rangle  = 0, \quad
 \left \langle \Omega^{\textrm{FP}(1)}_{i}, \frac{m_{i}c_{i}^{2}}{2} \right \rangle  = 0,
\end{split}
\end{align}
then as per Eq.(\ref{componentrelax1}) equilibrium $u_{i\alpha} = u_{\alpha}$ and $T_{i} = T$, therefore $\Omega_{i}^{\rm FP(L)} = 0$ then reduces to

\begin{equation}
\label{mb1}
\partial_{c_{i\alpha}}\left((c_{i\alpha} - u_{\alpha})f_{i} + \frac{k_{B}T}{m_{i}}\frac{\partial f_{i}}{\partial c_{i\alpha}}\right) = 0.
\end{equation}
Integrating Eq.(\ref{mb1}) with respect to the velocity space and using the fact that the distribution function and its derivatives tend to zero at infinity. We have

\begin{equation}
\label{mb2}
(c_{i\alpha} - u_{\alpha})f_{i} + \frac{k_{B}T}{m_{i}}\frac{\partial f_{i}}{\partial c_{i\alpha}} = 0.
\end{equation}
Solving Eq.(\ref{mb2}), we get the Maxwell-Boltzmann distribution as the solution. To find the equilibrium distribution   function for the high Schmidt case, we first note that

\begin{equation}
 \sum_{i}\left\langle \Omega^{\textrm{FP}(H)}_{i}, m_{i}\xi_{i\alpha}\xi_{i\beta} \right\rangle = \frac{2}{\tau_{2}} \left(\frac{k_BT}{m_i} - \frac{\theta_{\alpha\beta}}{m_i}\right),
\end{equation}
which suggests that $\Omega^{\textrm{FP}(H)} = 0 \implies \theta_{\alpha\beta}/m_i = (k_BT/m_i)\delta_{\alpha\beta}$. Hence, $\Omega^{\textrm{FP}(H)} = 0$ then reduces to Eq.(\ref{mb1}), the solution for which is the Maxwell-Boltzmann distribution.

Furthermore, the model must be consistent with the indifferentiability principle, i.e, one must be able to recover the Fokker-Planck approximation for single component case. In the case where $\tau_1 = \tau_2 = \tau$ and $m_A = m_B = m$, the Fokker-Planck collision kernel for binary mixtures outlined in Eq.(\ref{logFP}) reduces to the approximation for single component case indicating that proposed model abides by indifferentiability principle.

As demonstrated above, the proposed model does indeed satisfy the conservation laws, $H-$theorem, zero of collision and indifferentiability principle. Thus, this model is an acceptable approximation to the Boltzmann equation for binary mixtures.

\section{Transport coefficients}

In order to obtain the transport coefficients, we perform the Chapman-Enskog expansion, wherein the time derivative, distribution function and other relevant variables are represented as a series with $\textrm{Kn}$ acting as the smallness parameter \citep{chapman1970mathematical}. The distribution function is expressed in series form as

\begin{align}
 \begin{split}
 f_{i} &= f^{\textrm{MB}}_{i} + \textrm{Kn} f_{i}^{(1)} + \textrm{Kn}^2 f_{i}^{(2)} + ...,
 \end{split}
\end{align}
with the following constraints imposed on $f_{i}$
\begin{align}
\begin{split}
\label{ceConstrains}
 \langle f_{i}^{(n)}, \{ m_{i}, \sum_{i=A,B} m_{i}c_{i\alpha}, \sum_{i=A,B} m_{i}c_{i}^2/2 \} \rangle &= \{0, \textbf{0}, 0\}, \forall \quad n \geq 1 \\
\end{split}
\end{align}
These constraints ensure that component density, mixture momentum and energy are slow moments. The higher order moments in series form are
\begin{align}
\begin{split}
 \sigma_{\alpha\beta} &= \textrm{Kn} \sigma _{\alpha\beta} ^{(1)} + \textrm{Kn}^2 \sigma_{\alpha\beta} ^{(2)} + ..., \\
 q_{\alpha} &= \textrm{Kn} q_{\alpha} ^{(1)} + \textrm{Kn}^2 q_{\alpha} ^{(2)} + ...,
 \end{split}
\end{align}
as the stress and heat flux are zero at equilibrium and expected to be a function of slow moments otherwise. The time derivative is expressed in series form as
\begin{align}
 \begin{split}
 \partial _t &= \partial _t^{(0)} + \textrm{Kn} \partial_t ^{(1)} + \textrm{Kn}^2 \partial_t ^{(2)}+ .... 
 \end{split}
\end{align}

The time derivative of $f_i$  at zeroth order is computed using \citep{liboff2003kinetic}
\begin{equation}
 \partial_{t}^{(0)} f_i^{\rm MB}(\rho_i, \textbf{u}, T) = \frac{\partial f_i^{\rm MB}}{\partial \rho_i}\cdot \partial_{t}^{(0)} \rho_i + \frac{\partial f_i^{\rm MB}}{\partial \textbf{u}} \cdot \partial_{t}^{(0)} \textbf{u} + \frac{\partial f_i^{\rm MB}}{\partial T} \cdot \partial_{t}^{(0)} T,
\end{equation}
where the expression for time derivatives of the conserved variables can be calculated from the conservation laws mentioned in Eq.(\ref{eq:conLaw})

In order to find an expression for the viscosity, we first calculate the stress evolution equation. For the first model it has the form

\begin{align}
\begin{split}
\partial_{t}\sigma_{\alpha \beta} + \partial_{\gamma} (\sigma_{\alpha \beta} u_{\gamma})+ 2 p \overline{\partial_{\alpha} u_{\beta}}+2\overline{\sigma_{\alpha\gamma}\partial_{\gamma}  u_{\beta}}+\partial_{\gamma}Q_{\alpha \beta \gamma}+ \frac{4}{D+2} \overline{\partial_{\alpha}q_{\beta}} = \\
- \frac{2}{\tau_{1}} \left(\sigma_{\alpha \beta} +  \rho \overline{u_{\alpha} u_{\beta}} - \sum_{i=A,B}\rho_{i}\overline{ u_{i\alpha}u_{i\beta}}\right) - \frac{2}{\tau_{2}}
\left( \sum_{i=A,B} \rho_{i} \overline{ u_{i\alpha} u_{i\beta}} - \rho \overline{ u_{\alpha} u_{\beta}}\right),
\end{split}
\end{align}
where $Q_{\alpha\beta\gamma} = \sum_{i=A,B} \langle m_{i}\overline{\xi_{i\alpha}\xi_{i\beta}\xi_{i\gamma}} \rangle$. Retaining terms upto $\mathcal{O}(\textrm{Kn})$, the stress evolution equation yields

\begin{equation}
2 p \overline{\partial_{\alpha}u_{\beta}} = -\frac{2\sigma_{\alpha\beta}^{(1)}}{\tau_{1}}.
\end{equation}
and comparing with the Navier-Stokes law for stress tensor, we have

\begin{equation}
\mu = \frac{p\tau_1}{2}.
\end{equation}
Similarly, for the second model the right hand side of the stress evolution equation is

\begin{equation}
\sum_{i=A,B} \langle m_{i}\overline{\xi_{i\alpha}\xi_{i\beta}}, \Omega^{\textrm{FP}(2)}_{i} \rangle = - \frac{2}{\tau_{2}} \sigma_{\alpha \beta}.
\end{equation}
Hence, the expression for viscosity for this model is

\begin{equation}
\mu = \frac{p \tau_{2}}{2}.
\end{equation}

Similarly, the expression for diffusion coefficient can be calculated by considering the relaxation of diffusion flux defined as

\begin{equation}
 V_{\alpha} = m_{AB}(u_{A\alpha} - u_{B\alpha}),
\end{equation}
where $m_{AB} = (\rho_A \rho_B)/\rho$. Diffusive flux essentially quantifies the difference between the momentum of a given component and the momentum of the mixture. The series expansion for this quantity is


\begin{equation}
 V_{\alpha} = \textrm{Kn} V^{(1)}_{\alpha} + \textrm{Kn}^{2} V^{(2)}_{\alpha} + ...
\end{equation}
Similar to stress and heat flux, at equilibrium the diffusive flux attains zero values as momenta of both components relax to the mixture momentum. In order to calculate the expression for $V_{\alpha}$, we write the expression for individual component velocities. For the first model, we have

\begin{align}
 \begin{split}
 \partial _{t} \rho_{A} u_{A\alpha} + \partial _{\alpha} P_{A\alpha\beta} &= \frac{1}{\tau_2}(\rho u_{\alpha} - \rho_{A}u_{A\alpha}),\\
 \partial _{t} \rho_{B} u_{B\alpha} + \partial _{\alpha} P_{B\alpha\beta} &= \frac{1}{\tau_2}(\rho u_{\alpha} - \rho_{B}u_{B\alpha}),
 \end{split}
\end{align}
where $P_{i\alpha\beta} = \langle m_{i}c_{i\beta}c_{i\beta} \rangle$ and at equilibrium attains the value $P_{i\alpha\beta} = p_{i}\delta_{\alpha\beta} + \rho_{i}u_{\alpha}u_{\beta}$. After subtracting one equation from another and collecting terms upto $\mathcal{O}(Kn)$, we have

\begin{align}
\label{diffusionCoeff}
\begin{split}
 \partial_t^{(0)} (\rho_A - \rho_B) u_{\alpha} + \partial_{\beta} \left[(n_A - n_B)k_BT_0 \delta_{\alpha\beta} + (\rho_A - \rho_B)u_{\alpha}u_{\beta} \right] = -\frac{2}{\tau_2}V_{\alpha\beta}^{(1)}
\end{split}
\end{align}
The temporal derivatives are replaced using

\begin{equation}
 \partial_{t}^{(0)} \rho_{i} = -\partial_{\alpha}(\rho_{i}u_{\alpha}), \quad \partial_{t}^{(0)} \rho u_{\alpha} = -\partial_{\beta} (nk_BT_0 \delta_{\alpha\beta} + \rho u_{\alpha}u_{\beta}).
\end{equation}
After some rearrangement Eq.(\ref{diffusionCoeff}) takes the form

\begin{align}
\label{XA}
\begin{split}
 V_{\alpha}^{(1)} &= \tau_2 \left[ Y_{A}\partial_{\alpha}p - p\partial_{\alpha}X_{A} - X_{A}\partial_{\alpha}p \right],
 \end{split}
\end{align}
where $X_i = n_i/n$ is the component mole fraction and $Y_i = \rho_{i}/\rho$ is the component mass fraction. Rearranging Eq.(\ref{XA}) we have

\begin{equation}
\label{fdv1}
\partial_{\alpha} X_{A} = - \frac{V_{\alpha}^{(1)}}{ \tau_{2}p} + (Y_{A} - X_{A}) \frac{\partial_{\alpha} p}{p}.
\end{equation}
This has the same form as the Stefan - Maxwell equation \citep{bergman2011fundamentals} which governs the diffusion in multicomponent systems, and for binary mixtures is

\begin{equation}
\label{stefan}
\partial_{\alpha} X_{A} = \frac{X_{A} X_{B}}{D_{AB}} \frac{V_{\alpha}}{m_{AB}} + (Y_{A} - X_{A}) \frac{\partial_{\alpha} p}{p}.
\end{equation}
Comparing Eq.(\ref{fdv1}) with the Stefan-Maxwell equation, we get the following expression for the diffusion coefficient.

\begin{equation}
D_{AB} = X_{A} X_{B} \frac{p}{m_{AB}} \tau_{2}.
\end{equation}
The Schmidt number can now be computed as

\begin{equation}
{\rm Sc} =  \frac{\mu}{\rho D_{AB}} = \frac{\tau_{1}}{2\tau_{2}} \frac{m_{AB}}{X_{A} X_{B}}\frac{1}{\rho} = \frac{\tau_{1}}{ 2\tau_{2}} \frac{Y_{A} Y_{B}}{X_{A} X_{B}}.
\end{equation}
Existence of $H-$ theorem for this model suggests that $\tau_1 \leq \tau_2$, hence

\begin{equation}
{\rm Sc} \leq \frac{Y_{A} Y_{B}}{2 X_{A} X_{B}}.
\end{equation}
This model has an upper limit on the Schmidt number and this is in accordance with the characteristics of the quasi-equilibrium distribution. Similarly, for the second model, the Schmidt number is calculated as

\begin{equation}
{\rm Sc} =  \frac{\mu}{\rho D_{AB}} = \frac{\tau_{2}}{2\tau_{1}} \frac{m_{AB}}{X_{A} X_{B}}\frac{1}{\rho} = \frac{\tau_{2}}{ 2\tau_{1}} \frac{Y_{A} Y_{B}}{X_{A} X_{B}}.
\end{equation}
and since the limitation $\tau_1 \leq \tau_2$ exists, as consistent with the hypothesis there is a lower bound on the Schmidt number, which is

\begin{equation}
{\rm Sc} \geq \frac{Y_{A} Y_{B}}{2 X_{A} X_{B}}.
\end{equation}
Hence, both models in conjunction can cover a large range of Schmidt numbers.

\section{\label{sec:nal} Numerical scheme}

A Fokker-Planck equation which describes the evolution of probability density function of the random variable $\mathbf{\eta}$, is of the form

\begin{equation}
 \frac{d\mathcal{P}(\mathbf{\eta}, t)}{dt} = -\Lambda^{(1)}_{\alpha}(\mathbf{\eta},t)\frac{\partial \mathcal{P}(\mathbf{\eta}, t)}{\partial \eta_{\alpha}} + \frac{\zeta^{(1)}_{\alpha\beta}(\mathbf{\eta},t)}{2}\frac{\partial^{2} \mathcal{P}(\mathbf{\eta}, t)}{\partial \eta_{\alpha}\partial \eta_{\beta}} - \Lambda^{(2)}_{\alpha}(\mathbf{\eta},t)\frac{\partial \mathcal{P}(\mathbf{\eta}, t)}{\partial \eta_{\alpha}} + \frac{\zeta^{(2)}_{\alpha\beta}}{2}(\mathbf{\eta},t)\frac{\partial^{2} \mathcal{P}(\mathbf{\eta}, t)}{\partial \eta_{\alpha}\partial \eta_{\beta}},
\end{equation}
where $\mathbf{\Lambda}^{(i)}$ are the drift terms and $\mathbf{\zeta}^{(i)}$ are the diffusion coefficients. This form of Fokker-Planck equation is equivalent to the Langevin equation \citep{risken1996fokker}

\begin{equation}
\label{langevin}
\dot{\eta}_{\alpha} = h^{(1)}_{\alpha} (\mathbf{\eta}, t) + g^{(1)}_{\alpha \beta} (\mathbf{\eta}, t) \Gamma_{\beta}(t) + h^{(2)}_{\alpha}(\mathbf{\eta}, t) + g^{(2)}_{\alpha\beta} (\mathbf{\eta}, t ) \Gamma_{\beta}^{\prime}(t),
\end{equation}
where $\textbf{h}^{(i)}$ are the the drift terms, $\textbf{g}^{(i)}$ the diffusion coefficients and $\mathbf{\Gamma}, \mathbf{\Gamma}^{\prime}$ are Gaussian distributed random numbers which hold the following properties

\begin{equation}
\langle \Gamma_{\alpha}(t) \rangle = 0, \quad \langle \Gamma_{\alpha}(t) \Gamma_{\beta}(t^{\prime}) \rangle = \delta(t-t^{\prime}) \delta_{\alpha\beta}.
\end{equation}
Under these conditions, the following relations hold \citep{risken1996fokker}

\begin{align}
\begin{split}
 \Lambda^{(1)}_{\alpha} &= h^{(1)}_{\alpha}(\mathbf{\eta}, t), \quad
 \zeta^{(1)}_{\alpha\beta} = g^{(1)}_{\alpha\gamma}g^{(1)}_{\gamma\beta}\\
 \Lambda^{(2)}_{\alpha} &= h^{(2)}_{\alpha}(\mathbf{\eta}, t), \quad
 \zeta^{(2)}_{\alpha\beta} = g^{(2)}_{\alpha\gamma}g^{(2)}_{\gamma\beta}
\end{split}
\end{align}
The central idea is that the solution to Fokker-Planck equation is approximated by considering an ensemble of trajectories generated by the Langevin dynamics. In this case, a large number of particles have their positions and velocities updated using Eq.(\ref{langevin}). We now discuss the numerical scheme for the two cases.

\subsection{Low Schmidt limit}

For the first model the equivalent Langevin equations are

\begin{align}
\begin{split}
\label{LE1}
\frac{{{\rm d}x}_{\alpha}}{{\rm d}t}  &= c_{i\alpha},\\
\frac{{\rm d}c_{i\alpha}}{{\rm d}t} &= - \left(\frac{1}{\tau_{\textrm{eff}}} \right) ( c_{i\alpha} - u_{i\alpha})  - \frac{1}{\tau_{2}}( c_{i\alpha} - u_{\alpha}) + \sqrt{\frac{2 k_{B} T_{i}}{m_{i}}} {\rm d}W_{\alpha} + \sqrt{\frac{2 k_{B} T}{m_{i}}} {\rm d}W_{\alpha}^{\prime},
\end{split}
\end{align}
where ${\rm d}W_{\alpha}$ and ${\rm d}W_{\alpha}^{\prime}$ denote random forces with following statistics

\begin{equation}
\langle{\rm d}W_{\alpha} \rangle = 0, \quad \langle{\rm d}W_{\alpha}^{\prime}\rangle = 0, \quad \langle{\rm d}W_{\alpha} {\rm d}W_{\alpha}^{\prime}\rangle  = 0.
\end{equation}
More specifically, ${\rm d}W = W(t+\Delta t) - W(t)$ is the standard Weiner process, where $W(t)$ is a rapidly changing random force with  mean and variance as \citep{gardiner1985stochastic}

\begin{equation}
\langle {\rm d} W_{\alpha}(t)\rangle = 0, ~~~~\langle {\rm d} W_{\alpha}  {\rm d}W_{\beta}\rangle = {\rm d}t \delta_{\alpha \beta}.
\end{equation}
Thus, the detailed binary collision description is approximated by a random collision with a heat bath in the model.

These Langevin equations can be solved efficiently using the the stochastic version of the Verlet algorithm. For the present model the discretization scheme we have used is \citep{kloeden2013numerical,singh2015fokker}

\begin{align}
\begin{split}
\label{SV1}
x^{(1)}_{\alpha} = x_{\alpha}(t) + \frac{1}{2} c_{i\alpha}(t) \Delta t,\\
c_{i\alpha}(t + \Delta t) = c_{i\alpha}(t) - \left(\frac{\vartheta_{1}}{1 + \vartheta_{1}/2}\right) (c_{i\alpha}(t) - u_{i\alpha}) - \left(\frac{\vartheta_{2}}{1 + \vartheta_{2}/2} \right) (c_{i\alpha}(t) - u_{\alpha})\\
+  \frac{\sqrt{2\mathcal{D}^{(1)}_{i}\vartheta_{1}}}{1 + \vartheta_{1}/2}\phi_{\alpha} +  \frac{\sqrt{2\mathcal{D}^{(2)}_{i}\vartheta_{2}}}{1 + \vartheta_{2}/2}\phi^{\prime}_{\alpha},\\
x_{\alpha}(t + \Delta t) = x_{\alpha}^{(1)} + \frac{1}{2} c_{i\alpha}(t + \Delta t) \Delta t,
\end{split}
\end{align}
where $\vartheta_{1} = \Delta t/\tau_{\textrm{eff}}$, $\vartheta_{2} = \Delta t/\tau_{2}$ and $\phi_{\alpha},\phi_{\alpha}^{\prime}$ are Gaussian random numbers with mean zero and variance unity, $\mathcal{D}^{(1)}_{i}$ and $\mathcal{D}^{(2)}_{i}$ are $k_BT_i/m_i$ and $k_BT/m_i$ respectively. The recently proposed ``Molecular Dice'' algorithm \citep{agrawal2018molecular} was used to generate these Gaussian random numbers, which indicated considerable increase in efficiency without any loss of accuracy. This scheme works efficiently for small time steps such that $\textrm{max}\{\vartheta_{1}, \vartheta_{2}\} \leq 0.001$.

\subsection{High Schmidt limit}

The formulation for this model remains largely unchanged and the equivalent Langevin equations are

\begin{align}
\begin{split}
\label{LE2}
\frac{{{\rm d}x}_{\alpha}}{{\rm d}t}  &= c_{i\alpha},\\
\frac{{\rm d}c_{i\alpha}}{{\rm d}t} &= - \left(\frac{1}{\tau_{\textrm{eff}}} \right) ( c_{i\alpha} - u_{\alpha})  - \frac{1}{\tau_{2}}( c_{i\alpha} - u_{\alpha}) + \sqrt{2} \theta^{\prime}_{i\alpha\beta} {\rm d}W_{\beta} + \sqrt{\frac{2 k_{B} T}{m_{i}}} {\rm d}W_{\alpha}^{\prime},
\end{split}
\end{align}
where $\theta^{\prime}_{i\alpha\gamma}\theta^{\prime}_{i\gamma\beta} = \theta_{i\alpha\beta}/m_{i}$ and $\theta^{\prime}_{\alpha\beta}$ can be calculated by using Cholesky decomposition of $\theta_{\alpha\beta}/m_{i}$. The discretization scheme for this model is

\begin{align}
\begin{split}
\label{SV2}
x^{(1)}_{\alpha} = x_{\alpha}(t) + \frac{1}{2} c_{i\alpha}(t) \Delta t,\\
c_{i\alpha}(t + \Delta t) = c_{i\alpha}(t) - \left(\frac{\vartheta_{1}}{1 + \vartheta_{1}/2}\right) (c_{i\alpha}(t) - u_{\alpha}) - \left(\frac{\vartheta_{2}}{1 + \vartheta_{2}/2} \right) (c_{i\alpha}(t) - u_{\alpha})\\
+  \frac{\sqrt{2\vartheta_{1}}\theta^{\prime}_{i\alpha\beta}}{1 + \vartheta_{1}/2}\phi_{\beta} +  \frac{\sqrt{2\mathcal{D}^{(2)}_{i}\vartheta_{2}}}{1 + \vartheta_{2}/2}\phi^{\prime}_{\alpha},\\
x_{\alpha}(t + \Delta t) = x_{\alpha}^{(1)} + \frac{1}{2} c_{i\alpha}(t + \Delta t) \Delta t.
\end{split}
\end{align}

\begin{figure}
	\centering
        \begin{subfigure}[b]{0.48\textwidth}
        \centering
          \includegraphics[width=\linewidth]{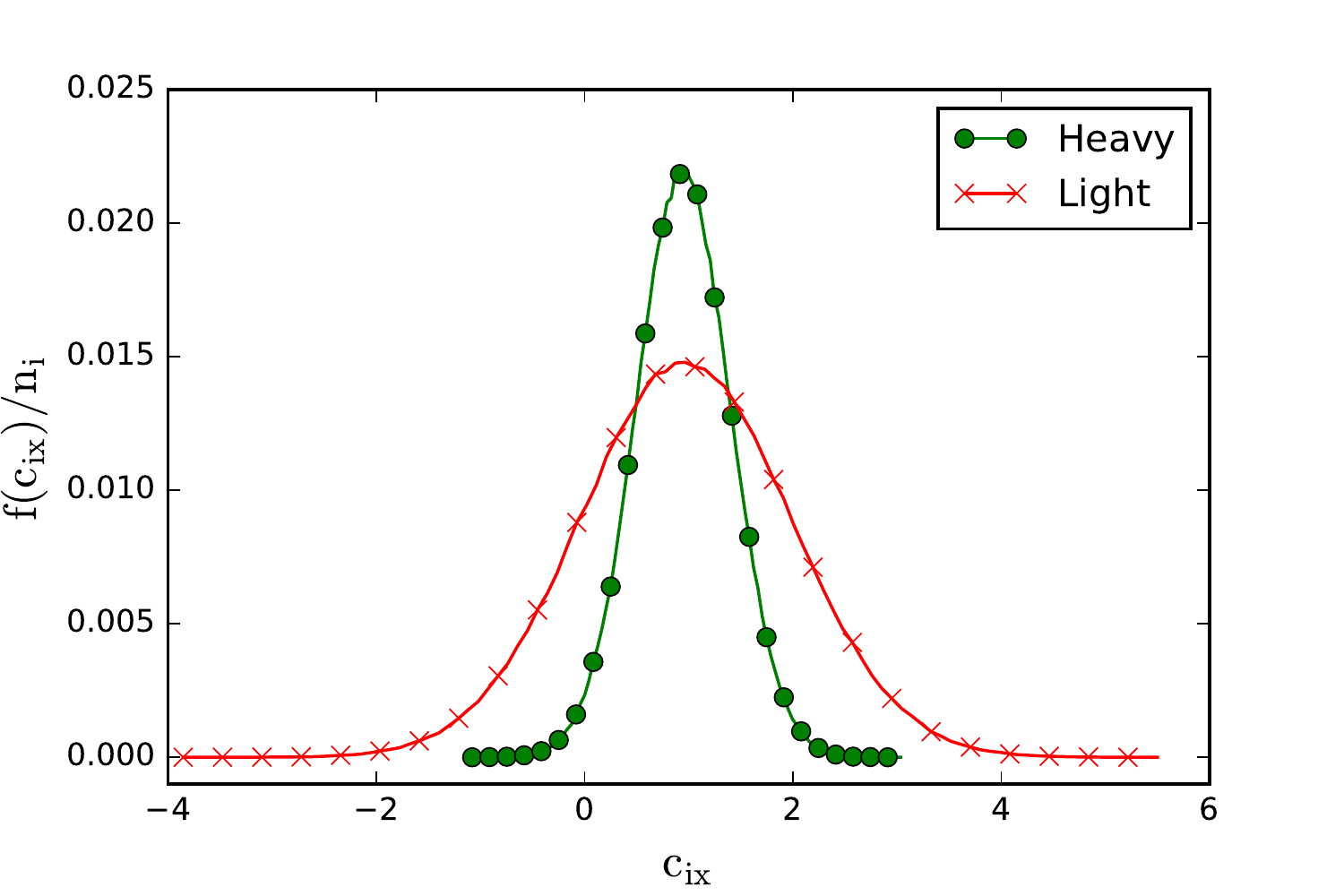}
          \caption{}
          \end{subfigure} \quad
        \begin{subfigure}[b]{0.48\textwidth}
        \centering
          \includegraphics[width=\linewidth]{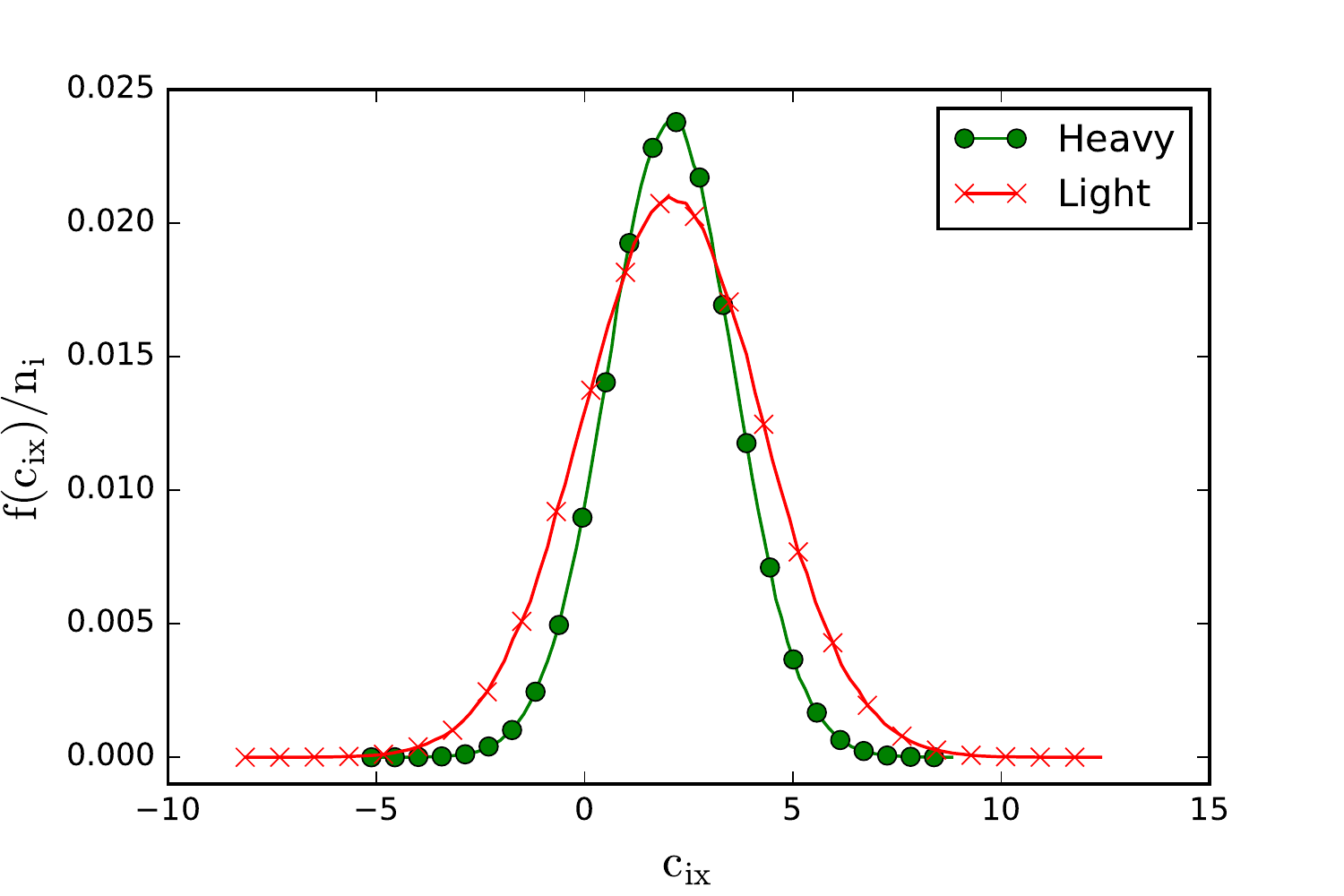}
            \caption{}
                \end{subfigure}   
         \caption{Plot of the distribution of velocities of the light and heavy component at equilibrium for a) Model I, and b) Model II.}  
         \label{distribution}    
  \end{figure}

\begin{figure}   
       \begin{subfigure}[b]{0.48\textwidth}
        \centering
          \includegraphics[width=2.75 in, height = 2.0 in]{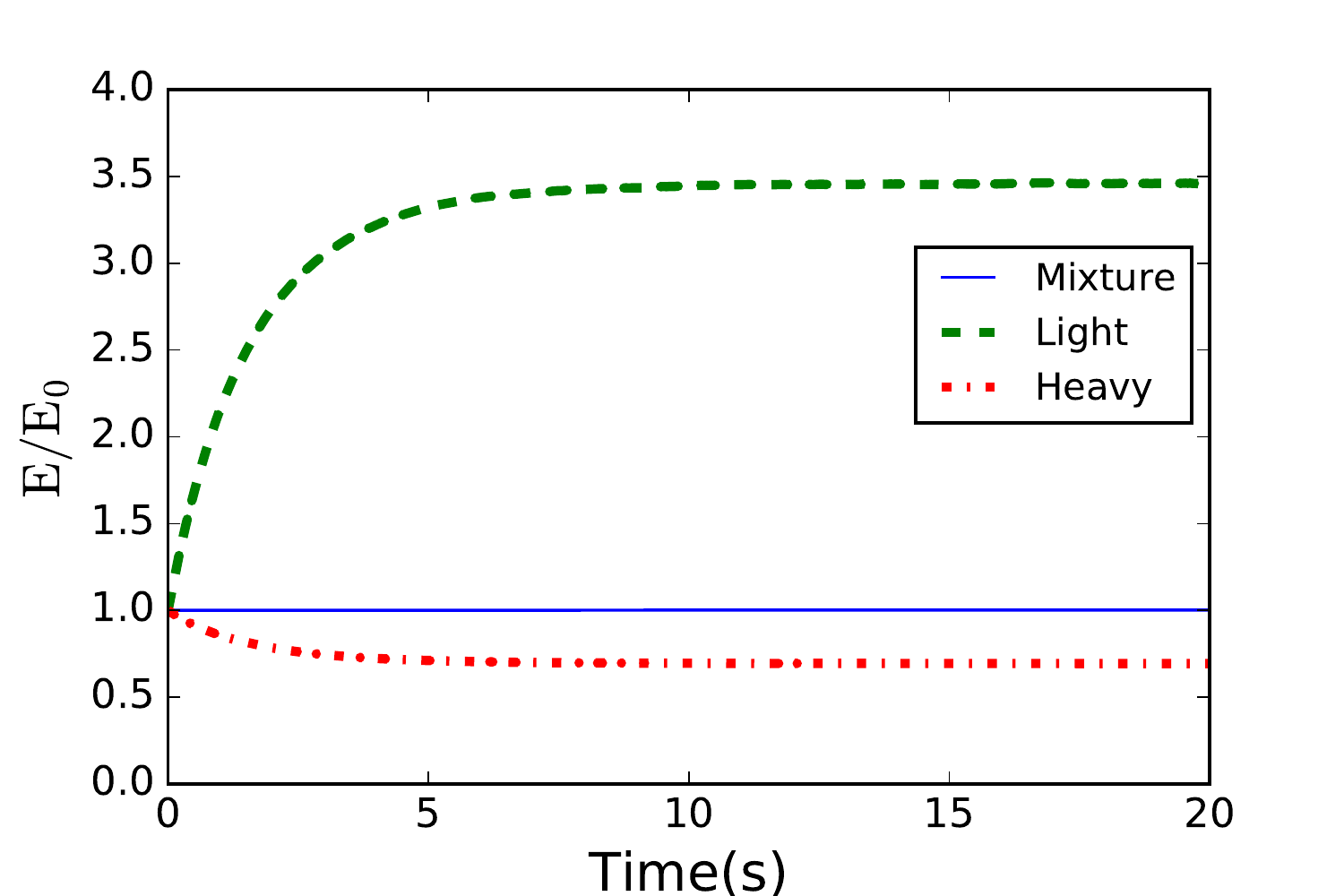}
          \caption{}
          \end{subfigure}
        \begin{subfigure}[b]{0.48\textwidth}
        \centering
          \includegraphics[width=2.75 in, height = 2.0 in]{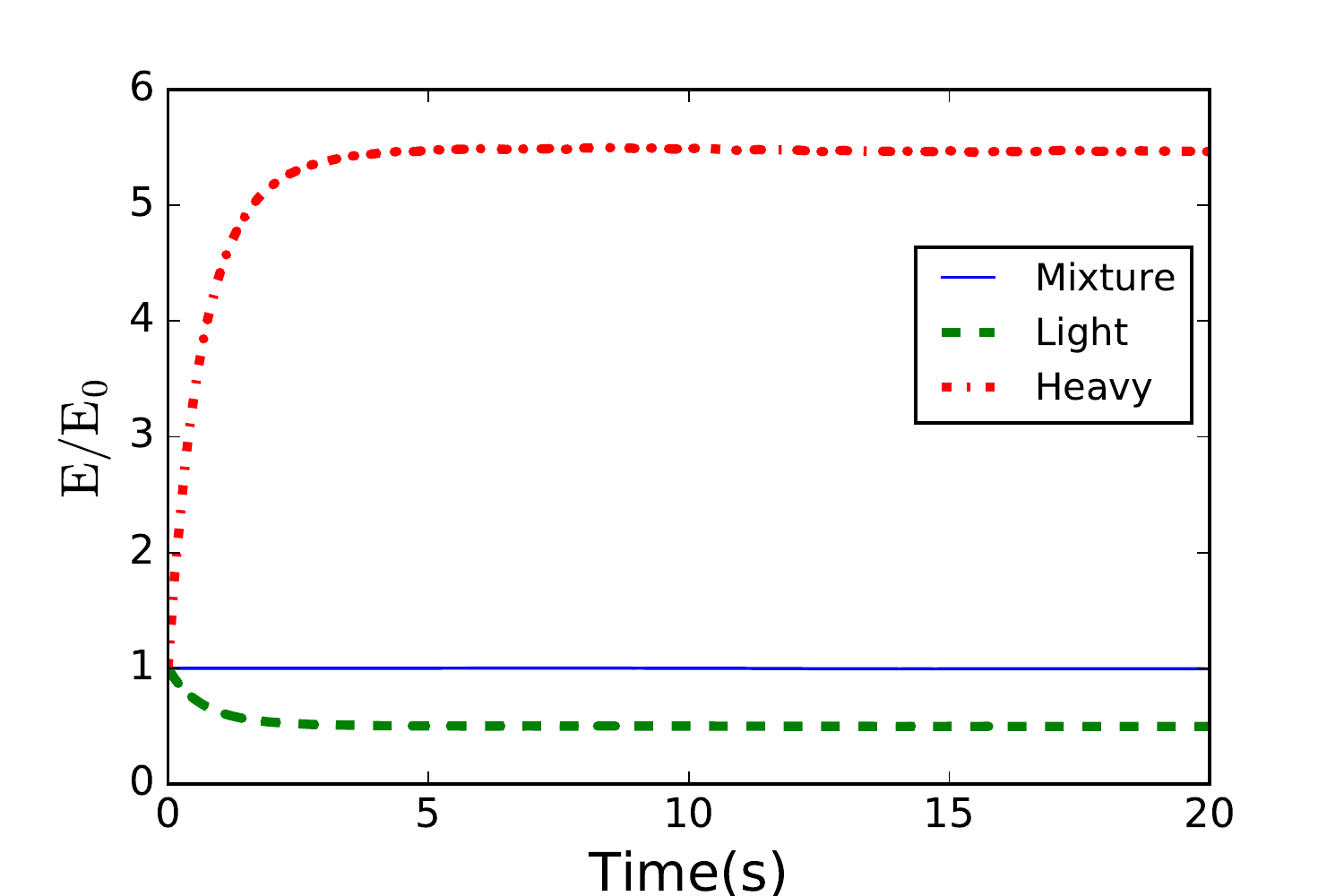}
          \caption{}
          \end{subfigure}
          \caption{Plot of ratio of energy at time $t$ to the initial energy ($E(t)/E_0$) vs. time for individual components and the mixture for a) Model I, and b) Model II.}         \label{energy}
        \end{figure}

In order to validate the numerical scheme, we started with a mixture with $m_B/m_A = 2$ with $N = 10^5$ particles in a single periodic box. For Model I, the velocities of the lighter particles were initialized uniformly in the range $[0,1)$ and the heavier particles in the range $[0,2)$. For Model II, the velocities of lighter particle were initialized with a Gaussian distribution with mean $4$ and variance $10$, and the heavier particles were Gaussian distributed with mean $1$ and variance $1$. The plots of energy of the two components and the mixture with time averaged over an ensemble of $15$ trajectories and the distribution of velocities in a particular direction, for both cases are shown in figure Eq.(\ref{distribution}) and Fig. \ref{energy}.
          
\section{Simulation results}

In this section, we present the results for three benchmark problems -- Graham's law for effusion, Couette flow and binary diffusion.

\subsection{Graham's law for effusion}

\begin{figure}
\centering
\includegraphics[scale=0.5]{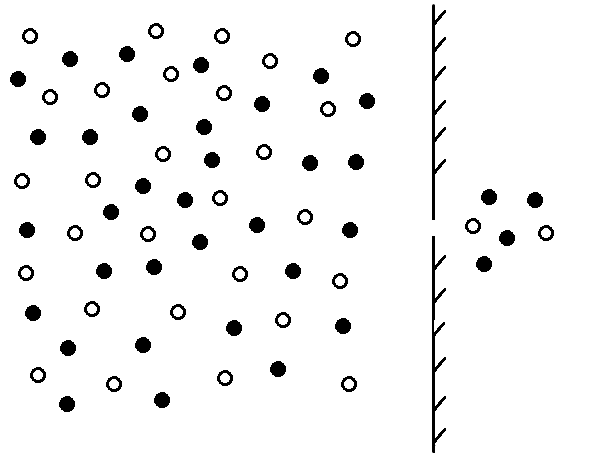}
\caption{The lighter particles (in this case filled) escape through the hole faster than the heavier particles, with a factor proportional to the square root of their mass ratios.}
\label{effusion}
\end{figure}

Effusion is a process wherein gas molecules escape through a small hole. The length parameter of this hole is much smaller than the mean free path of the gas, i.e, $d \ll \lambda_{\textrm{mfp}}$. A sketch of the process has been shown in Fig. \ref{effusion}. The number flux of the gas through this small hole is

\begin{figure}
\centering
\includegraphics[scale=0.7]{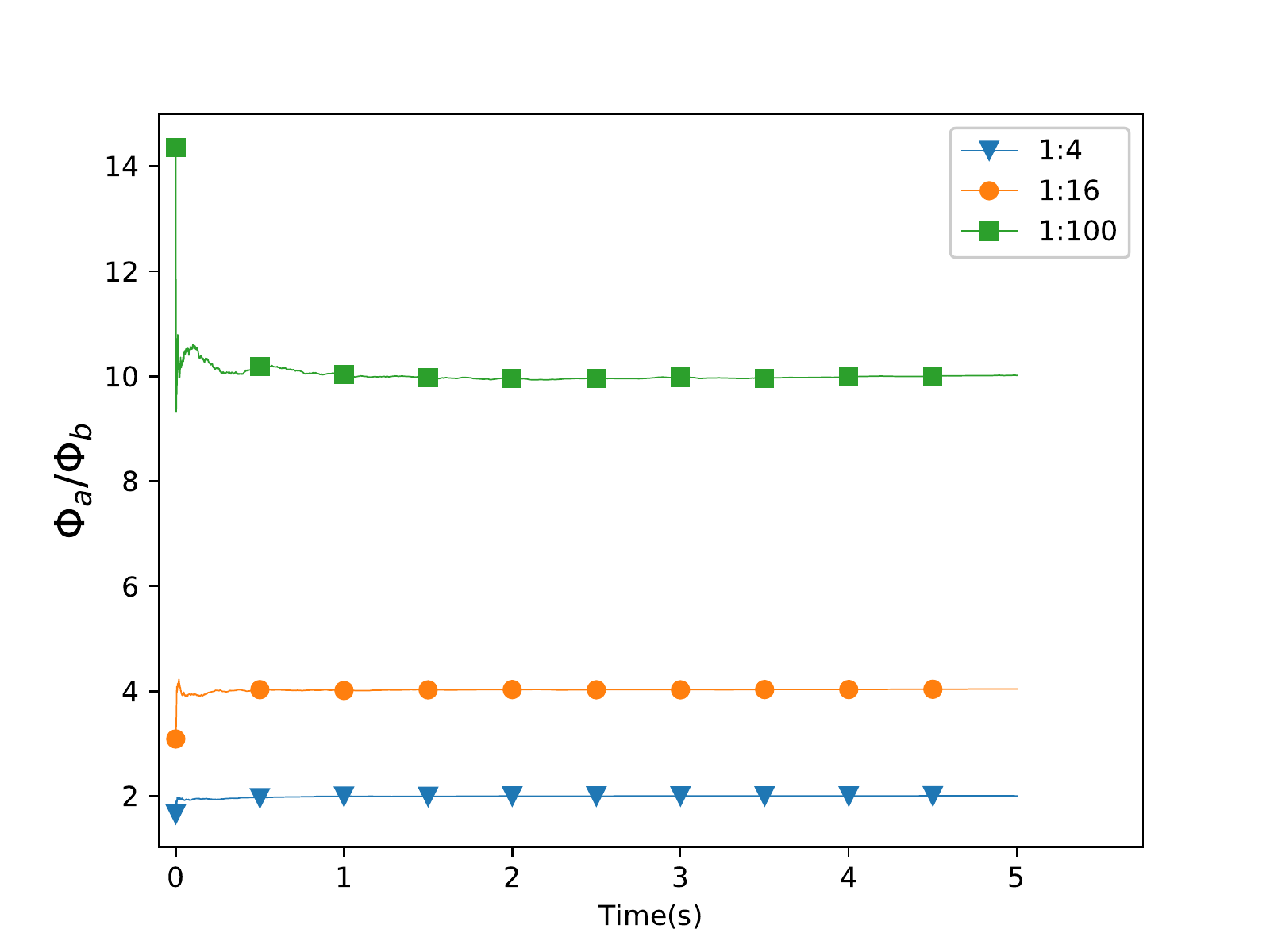}
\caption{Model I was used to simulate a setup that could mimic Graham's law for effusion. Plot shows that results observed are in great agreement with expected behaviour, for all three cases.}
\label{graham}
\end{figure}

\begin{equation}
\Phi_{i} = \langle c_{iz}, f(\textbf{c}_{i}) \rangle,
\end{equation}
where $\Phi_{i}$ is the number flux and $c_{iz}$ the molecular velocity in the direction perpendicular to the plane of the hole. By integrating over velocity space, facilitated by a shift to the spherical co-ordinate system, the expression of $\Phi_{i}$ is

\begin{equation}
 \Phi_{i} = \frac{P}{\sqrt{2\pi m_{i}k_BT}},
\end{equation}
where $P$ is the pressure and $T$ the temperature of the gas. Then, for a well-mixed binary mixture the ratio of the fluxes is \citep{mason1967graham}

\begin{equation}
 \frac{\Phi_{A}}{\Phi_{B}} = \sqrt{\frac{m_{B}}{m_{A}}}.
\end{equation}
We simulated this system for three mass ratios $m_{B}/m_{A} = 4$, $m_{B}/m_{A} = 16$ and $m_{B}/m_{A} = 100$. The boundary conditions in the transverse directions were taken to be periodic while maintaining constant pressure in the system. The results have been plotted in Fig. \ref{graham}. As can be seen, the simulations are in excellent agreement with the analytical solution.
\subsection{Couette Flow}

\begin{figure}
\centering
\includegraphics[scale=0.5]{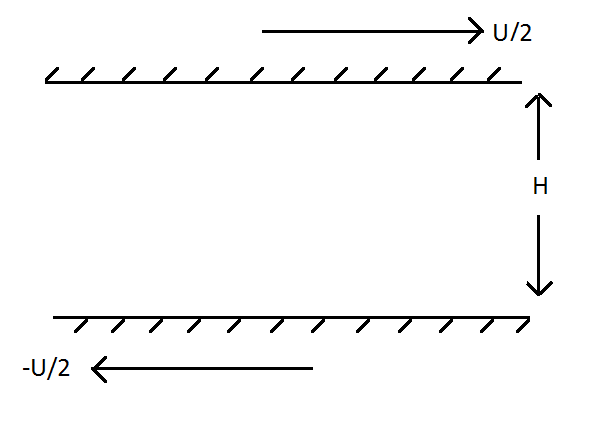}
\caption{A representative sketch of the Couette flow steup. Two walls with a separation $H$ are sheared in the opposite directions with velocity $U/2$.}
\label{couette}
\end{figure}

The setup of the problem is simple, fluid between two plates is sheared in opposite directions with equal magnitudes, a sketch of the problem is shown in Fig. \ref{couette}. In order to validate the model, we calculate the global stress tensor defined as \citep{SGK04}

\begin{equation}
 \Pi = -\frac{v_0}{2UP_0}P_{xy},
\end{equation}
where $P_0 = nk_BT_0$ is the reference pressure. This quantity is calculated in the entire range of rarefaction parameter, $\delta$, which is inverse of the Knudsen number and is defined as

\begin{equation}
 \delta = \frac{HP_0}{\mu v_0},
\end{equation}
where $\mu$ is the mixture viscosity and $v_0$ the characteristic molecular velocity of the mixture defined as

\begin{equation}
 v_0 = \sqrt{\frac{2k_BT_0}{m_0}},
\end{equation}
where $m_0 = C_0m_A + (1-C_0)m_B$, with $C_0$ being the concentration of the lighter component. We simulated the system for three mixtures Neon-Argon (Ne-Ar), Helium-Argon (He-Ar) and Helium-Xenon (He-Xe) for rarefaction parameters ranging from $[0.01, 40]$ for three different concentrations - $(0.1,0.5,0.9)$. The value for $\Pi$ was computed by averaging over $10^5$ iterations for each parameter and the results are tabulated in Table \ref{stress1}. The error bar (standard deviation) was of the same order for all parameters and ranged from $[0.00478, 0.00544]$. The results were found to be in good agreement with reported results \citep{SGK04}. This indicates that proposed method is indeed capable of simulating flows in a wide range of Knudsen numbers.
\addtolength{\tabcolsep}{2pt}
\begin{table}
\begin{center}
\begin{tabular}{c c c c c c c c c c}
& \multicolumn{3}{c}{Ne-Ar} & \multicolumn{3}{c}{He-Ar} &  \multicolumn{3}{c}{He-Xe} \\
    \hline
$\delta$ & $C_0$ = 0.1 & 0.5 & 0.9 & 0.1 & 0.5 & 0.9 & 0.1 & 0.5 & 0.9 \vspace{-0.2pt}\\
    \hline
0.01 & 0.27558 & 0.27266 & 0.27471 & 0.27004 & 0.24510 & 0.24381 & 0.26694 & 0.22559 & 0.19842\\
0.1 & 0.25295 & 0.25014 & 0.25216 & 0.24764 & 0.22383 & 0.22296 & 0.24442 & 0.20483 & 0.17994\\
1.0 & 0.16539 & 0.16324 & 0.16458 & 0.16141 & 0.14455 & 0.14650 & 0.15835 & 0.12959 & 0.11892\\
10.0 & 0.04141 & 0.04124 & 0.04159 & 0.04054 & 0.03886 & 0.04055 & 0.04091 & 0.03526 & 0.03706\\
40.0 & 0.01222 & 0.01219 & 0.01196 & 0.01220 & 0.01185 & 0.01217 & 0.01125 & 0.01165 & 0.01155\\
\end{tabular}
\end{center}
\caption{$\Pi$ values for Ne-Ar, He-Ar and He-Xe mixtures for three different compositions}
\label{stress1}
\end{table}

\subsection{Binary diffusion}

\begin{figure}
	\centering
        \begin{subfigure}[b]{0.48\textwidth}
        \centering
          \includegraphics[width=\linewidth]{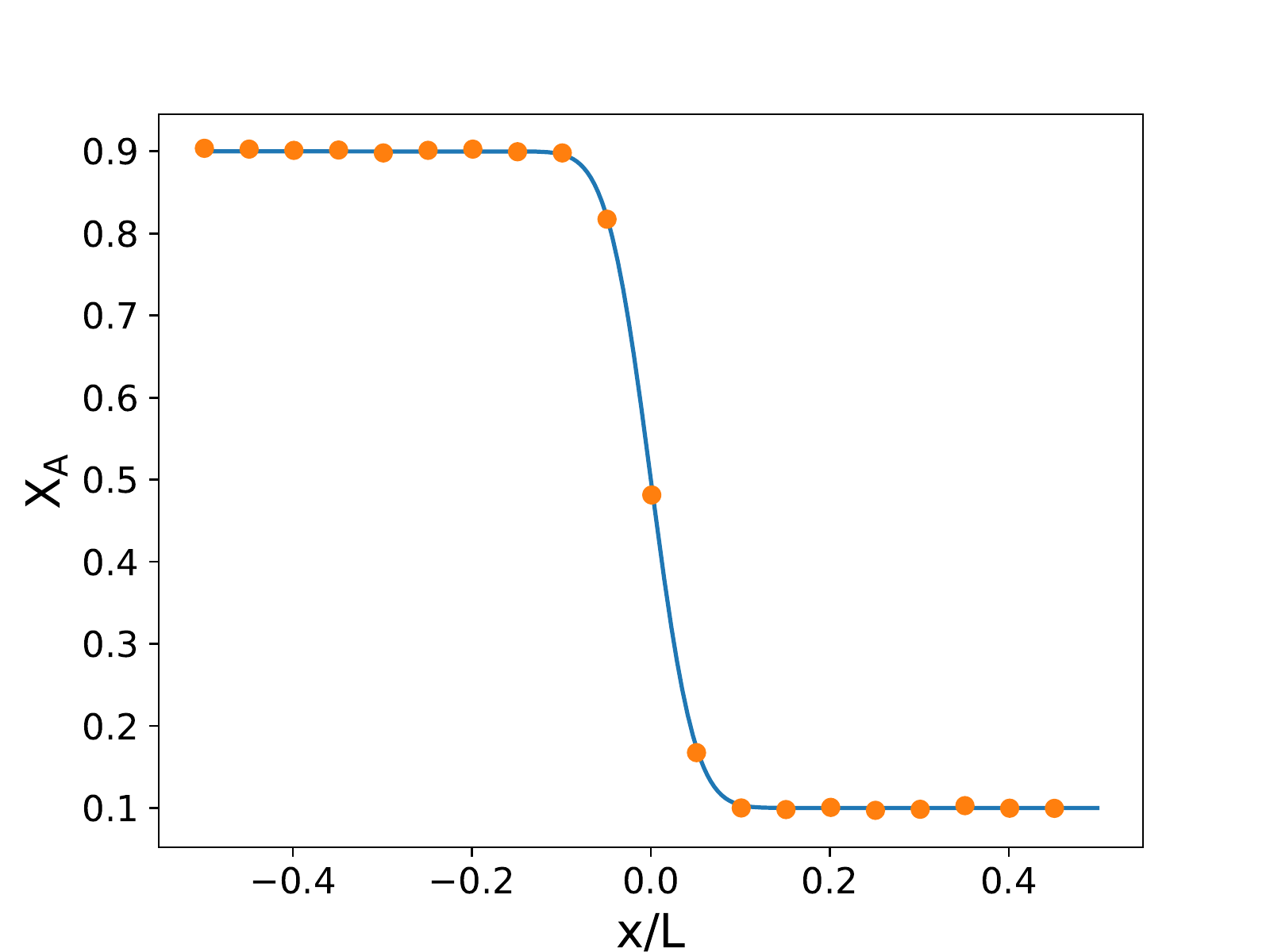}
          \caption{}
          \end{subfigure} \quad
        \begin{subfigure}[b]{0.48\textwidth}
          \includegraphics[width=\linewidth]{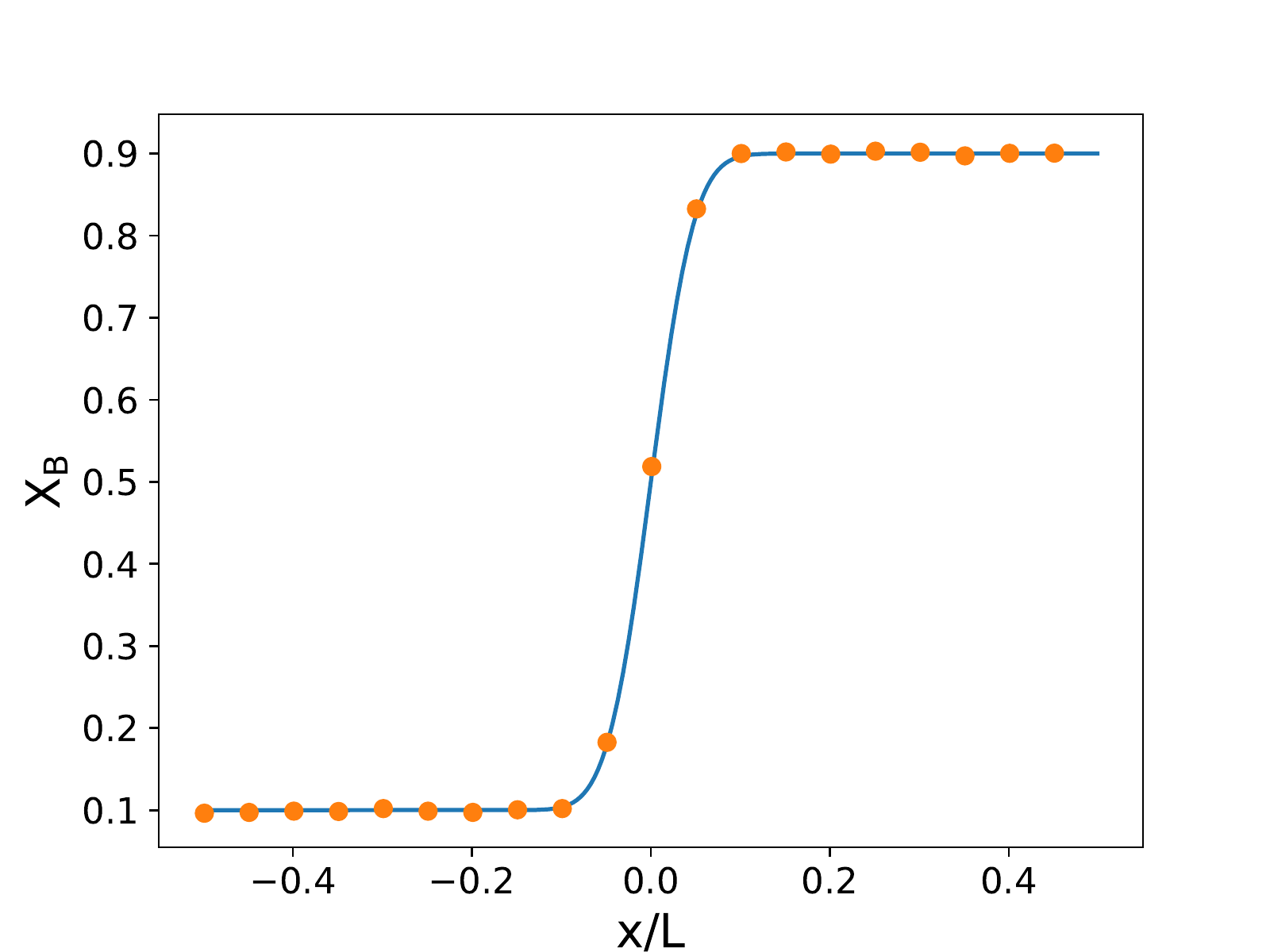}
          \caption{}
          \end{subfigure}
        \caption{Plot of the concentrations after 20,000 time steps of the component a) A and b) B, in comparison with the analytical solution given by Eq.(\ref{diffAnalytical}).}
          \label{static}
\end{figure}

The profile of the mixture in this setup is determined by the step function

\begin{align}
\begin{split}
X_A &= 90 \%, \quad X_B = 10\% \quad \textrm{if} \quad x < 0, \\
X_B &= 10 \%, \quad X_B = 90\% \quad \textrm{if} \quad x \geq 0,
\end{split}
\end{align}
where the mass ratio of the components was chosen to be $m_B/m_A = 5$. The step function is used instead of a smooth profile as it is a more severe check for the numerical scheme. Under the assumption that at infinity, the initial concentrations remains unchanged, this problems yields the analytical solution \citep{bergman2011fundamentals}

\begin{equation}\label{diffAnalytical}
 X_i = \left[\frac{1}{2} + \frac{\Delta X_i}{2} \textrm{erf} \left(\frac{x}{\sqrt{4D_{AB}t}} \right)\right],
\end{equation}
where $D_{AB}$ is the diffusion constant. The simulation was done for $20,000$ time steps and the plots for both the components compared against their respective analytical solutions are plotted in Fig. \ref{static}. The simulation results were very close to the analytical solution. This exercise proves that the value of $D_{AB}$ set by the numerical scheme is accurate.

\section{Outlook}
We developed a new thermodynamically consistent Fokker-Planck approximation to the Boltzmann equation for binary gas mixtures, based on quasi-equilibrium models. These models were subjected to numerical experiments like Graham's law, Couette flow and binary diffusion and it was determined that the algorithm is capable of simulating flow for a wide range of Knudsen numbers and diffusion coefficients. The extension of the existing Fokker-Planck model to binary mixtures, is an indication that it can be employed to solve for mixtures with many components. Future work is to extend this model to multi-component gas mixtures.

\acknowledgements{\textbf{Acknowlegements}}\\

S. K. Singh expresses his sincere thanks to the Department of Science and Technology (DST), SERB, India, for financial
support under Young Scientist Scheme (S. No. YSS/2015/002085).  S. Ansumali expresses his sincere thanks to Department of
Science and Technology, India for financial assistance (S.No.: SB2/S2/CMP-056/2013).

\section*{References}
\bibliography{binaryFP}
\end{document}